\documentclass[a4paper]{spie}  

\usepackage{authblk}
\usepackage{amsmath,amsfonts,amssymb}
\usepackage{graphicx}
\usepackage[colorlinks=true, allcolors=blue]{hyperref}

\title{A low-cost ultraviolet-to-infrared absolute quantum efficiency characterization system of detectors}

\newcommand\uoftastro{David A. Dunlap Dept. of Astronomy and Astrophysics, University of Toronto, 50 St. George Street, Toronto, ON, Canada M5S 3H4}
\newcommand\dunlap{Dunlap Institute for Astronomy and Astrophysics, University of Toronto, 50 St. George Street, Toronto, ON, Canada M5S 3H4}
\newcommand\uoftphysics{Department of Physics, University of Toronto, 60 St. George Street, Toronto, ON, Canada M5R 2M8}
\author[a,b]{Ajay S. Gill}
\author[b,c]{Mohamed M. Shaaban}
\author[a]{Aaron Tohuvavohu}
\author[a,b]{Suresh Sivanandam}
\author[a,b]{Roberto G. Abraham}
\author[a,b]{Seery Chen}
\author[a,b]{Maria R. Drout}
\author[a,b]{Deborah Lokhorst}
\author[a]{Christopher D. Matzner}
\author[a]{Stefan W. Mochnacki}
\author[a,b,c]{Calvin B. Netterfield}

\affil[a]{\uoftastro{}}
\affil[b]{\dunlap{}}
\affil[c]{\uoftphysics{}}

\authorinfo{E-mail: ajay.gill@mail.utoronto.ca}

\begin{document} 
\maketitle

\begin{abstract}
We present a low-cost ultraviolet to infrared absolute quantum efficiency detector characterization system developed using commercial off-the-shelf components. The key components of the experiment include a light source, a regulated power supply, a monochromator, an integrating sphere, and a calibrated photodiode. 
We provide a step-by-step procedure to construct the photon and quantum efficiency transfer curves of imaging sensors. We present  results for the GSENSE 2020 BSI CMOS sensor and the Sony IMX 455 BSI CMOS sensor. As a reference for  similar characterizations, we provide a list of parts and associated costs along with images of our setup. 
\end{abstract}


\section{INTRODUCTION}
\label{sec:intro}  


The ability of a sensor to convert incident photons into electrons (as a function of photon wavelength) is set by its quantum efficiency (QE). For a source with a given incoming photon flux, a sensor with a high QE will generate a higher number of electrons per second compared to a sensor with a low QE, allowing the former sensor to detect the source with a higher signal-to-noise ratio, given as

\begin{equation}
    \frac{\mathrm{S}}{\mathrm{N}} = \frac{N_{*}}{\sqrt{N_{*} + n_{\mathrm{pix}}(N_{\mathrm{S}} + N_{\mathrm{D}} + N_{\mathrm{R}}^{2}})} \propto N_{*}^{1/2}
\end{equation}

\noindent where $N_{*}$ is the number of electrons per second from the source (the \textit{signal}). The \textit{noise} terms are the shot noise from the source ($N_{*}^{1/2}$), number of electrons per second per pixel from the background ($N_{\rm S}$), number of thermally generated electrons per second per pixel ($N_{\rm D}$), and electrons per pixel due to read noise ($N_{\rm R}^{2}$). The photons from the source induce electrons in the valence band to jump into the conduction band of the semiconductor, where the electrons can freely move and be measured with readout electronics. The QE of a sensor can thereby set the sensitivity limit for the detection of astronomical sources. Understanding the QE of the sensor for an astronomical instrument is therefore crucial, as it informs the instrument design, the observation strategy, and the scientific capability of the instrument. 

In this paper, we present a low-cost detector characterization system built using commercial-off-the-shelf (COTS) components. First, we provide some references to literature on other QE measurement setups. Jacquot et al. (2011) \cite{Jacquot_2011} present their system for ultraviolet (UV) to near-infrared (NIR) QE measurements in vacuum. Sperlich and Stolz (2013)\cite{Sperlich_2013} present their QE measurement setup and results for one front-illuminated electron multiplying CCD (EMCCD) and five back-illuminated EMCCDs. Coles et al. (2017)\cite{Coles_2017} present their QE measurement system for the CCD sensors for the Vera-Rubin Observatory. Krishnamurthy et al. (2019) \cite{Krishnamurthy_2017} present their QE measurement system developed for the Transiting Exoplanet Survey Satellite (TESS) CCD detectors. Bastian-Querner et al. (2021)\cite{Bastian_Querner_2021} present their sensor characterization system for the ULTRASAT space telescope.   

In Section \ref{sec:theory}, we discuss the theory behind the photon transfer and QE transfer curves. In Section \ref{sec:experiment}, we present the experimental setup, the list of components, and the experimental procedure for constructing the photon and QE transfer curves. In Section \ref{sec:results}, we present results of the photon and QE transfer curves for two COTS sensors, the GSENSE 2020 BSI CMOS sensor (using the Aluma 2020 BSI camera from the vendor Diffraction Limited\footnote{https://diffractionlimited.com} and the Sony IMX 455 BSI CMOS sensor using the QHY600 camera from the vendor QHYCCD\footnote{https://www.qhyccd.com}). In particular, the IMX 455 CMOS sensor is popular for both amateur and professional astronomers alike (to be used for instance for the Argus Array\cite{Law_2022} and the SuperBIT\cite{Romualdez_2020} balloon-borne telescope). 

\section{Theory}
We briefly discuss the theory behind the photon transfer curve and the quantum efficiency transfer curves. For further details, we refer the reader to Janesick (2001) \cite{s_ccd} and Janesick (2007) \cite{s_ccd_2}.

\label{sec:theory}
\subsection{Photon transfer curve}

\begin{figure}[htb!]
\begin{center}
\begin{tabular}{c} 
\includegraphics[height=4cm]{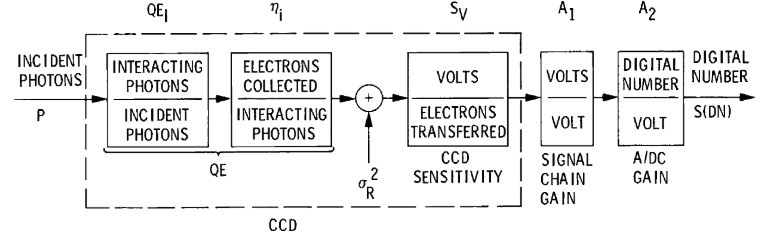}
\end{tabular}
\end{center}
\caption[example]
{\label{fig:block_qe}
  {Block diagram of a typical CCD with individual transfer functions\cite{s_ccd}. }}
\end{figure}

The block diagram of a typical sensor showing the individual transfer functions is shown in Figure \ref{fig:block_qe}. 

\noindent Given incident photons, the output signal of a CCD, $S$(ADU) for a given exposure time is given by

\begin{equation}
S[\mathrm{ADU}] = P \times \mathrm{QE_{I}} \times \eta_{i} \times S_{\rm{V}} \times
A_{\rm{CCD}} \times A_{1} \times A_{2}
\end{equation}

\begin{itemize}
\item S[ADU] is the average signal for a group of pixels [ADU], where ADU stands for analog-to-digital unit
\item $P$  is the average number of incident photons per pixel [photons/pixel]
\item QE$_{\rm{I}}$ is the interacting quantum efficiency [interacting photons/incident photons]
\item $\eta_{i}$ is the quantum yield [number of electrons generated, collected, and transferred per interacting photon]
\item $S_{\rm{V}}$ is the sensitivity of the sense node [V/electron]
\item $A_{\rm{CCD}}$ is the output amplifier gain [V/V]
\item $A_{1}$ is the gain of the signal processor [V/V]
\item $A_{2}$ is the gain of the ADC [ADU/V]
\item QE is the quantum efficiency of the sensor [numbers of electrons generated, collected, and transferred per incident photon]
\end{itemize}

\noindent The camera gain constant, $K$ [electron/ADU], is given as

\begin{equation}
K = \frac{1}{S_{\rm{V}} \times  A_{\rm{CCD}} \times A_{1} \times A_{2}}
\end{equation}

\noindent The photon transfer curve (PTC) is a powerful technique to characterize important parameters of a sensor, including the read noise, full well capacity, linearity, fixed pattern noise (FPN), dynamic range, and the gain. The PTC is generated by illuminating the sensor with a monochromatic uniform light source as a function of exposure time, starting with a dark frame and increasing the exposure time until saturation. After the bias offset has been subtracted from the signal, the noise is plotted as a function of signal for different exposure times in ADUs. The total noise includes read noise, shot noise, and FPN, all of which are added in quadrature.

\begin{equation}
\sigma_{\rm{TOTAL}}[\rm{ADU}] = \sqrt{\sigma_{\rm{READ}}[ADU]^{2} + \sigma_{\rm{SHOT}}[ADU]^{2} + \sigma_{\rm{FPN}}[ADU]^{2}} 
\end{equation}

\noindent The shot noise is given by

\begin{equation}
\sigma_{\rm{SHOT}} [\mathrm{ADU}] = \Bigg(\frac{S \, [\mathrm{ADU}]}{K\, \mathrm{[e^{-}/ADU]}}\Bigg)^{1/2}
\label{eqn:shot_noise}
\end{equation}

\noindent The FPN is given by

\begin{equation}
\sigma_{\mathrm{FPN}} = P_{\mathrm{N}} \times S[\mathrm{ADU}]
\label{eqn:fpn}
\end{equation}

\noindent where $P_{\mathrm{N}}$ is the fixed pattern noise quality factor (usually $\sim 1\%$ for CCD/CMOS sensors). To estimate the read noise and the camera gain $K$, the FPN must first be removed to obtain a shot plus read noise only curve. The FPN can be removed by pixel-by-pixel subtraction of two identical frames taken one after the other at the same exposure level. The resulting difference frame contains read and shot noise only, and a separate curve can be plotted. The camera gain constant $K$ [e$^{-}$/ADU] can then be estimated by fitting the shot plus read noise curve with a line of slope 1/2. Similarly, $P_{\mathrm{N}}$ can be estimated by fitting the total noise PTC with a slope of 1. The read noise in electrons can be estimating by finding the offset of the zero slope line and multiplying by $K$. The full well capacity in electrons can be estimated from the total noise PTC as well. The dynamic range of the sensor can also be estimated by the ratio of the full well capacity and the read noise.  

\subsection{Quantum efficiency transfer curve}
The quantum efficiency of a sensor is given as QE = QE$_{\rm{I}}$ $\times \eta_{i}$, where QE$_{\mathrm{I}}$ is the interacting QE [interacting photon per incident photon] and $\eta_{i}$ is the quantum yield [electroncs collected per interacting photon]. The electromagnetic power for a given incident photon rate on the sensor is

\begin{equation}
P = \frac{\text{incident photons}}{\text{s}} \times h\nu \;\; \Bigg[\frac{\text{erg}}{\text{s}} \Bigg]
\end{equation}

\noindent The electrons collected per second are

\begin{equation}
{\frac{\text{electrons collected}}{\text{s}}} = \text{QE} \times \frac{\text{incident photons}}{\text{s}} = \text{QE} \times \frac{P}{h\nu}
\end{equation}

\noindent The QE is measured by comparing the electrons collected on the sensor and a calibrated photodiode given the same input flux from a lamp with a regulated power supply. The photocurrent induced by the photodiode is 

\begin{equation}
I_{\mathrm{D}} = \frac{q \times \text{QE} \times P}{h \nu} = \frac{q \times \lambda \times \text{QE} \times P}{hc} \;\;\; [\text{Amperes}]
\end{equation}

\noindent The responsivity of a calibrated photodiode is typically provided in units of [A/W].

\begin{equation}
R_{\lambda} = \frac{I_{D}}{P} = \frac{q \times \lambda \times \text{QE}_{\mathrm{D}}}{hc} \;\;\; \Bigg[\frac{\text{A}}{\text{W}} \Bigg]
\end{equation}

\noindent The QE of the photodiode is then

\begin{equation}
\text{QE}_{\mathrm{D}} = \frac{R_{\lambda}}{\lambda} \times \frac{hc}{q} = 1239.842 \times \frac{R_{\lambda}}{\lambda}
\end{equation}

\noindent where $\lambda$ is in units of nanometers. The sensor count rate is

\begin{equation}
S_{\mathrm{CCD}} = \frac{S_{\mathrm{ADU}} \times K}{t_{\mathrm{exp}}} \;\;\; [\text{electrons/sec/pixel}]
\end{equation}

\noindent The QE of the sensor is then

\begin{equation}
\text{QE}_{\mathrm{sensor}} = \frac{\text{Sensor term}}{\text{Photodiode term}}
\label{eqn:qe_eqn}
\end{equation}

\begin{equation}
\text{QE}_{\text{sensor}} = \Bigg[\frac{S_{\mathrm{ADU}} \times K}{t_{\rm exp} \times A_{\text{sensor}}} \Bigg]  \div \Bigg[\frac{\lambda \times I_{\mathrm{D}}}{1239.842 \times q \times R_{\lambda} \times A_{\text{diode}}}\Bigg]
\label{eqn:qe}
\end{equation}

\noindent where $A_{\mathrm{diode}}$ is the active area of the photodiode [cm$^{2}$], and $A_{\mathrm{sensor}}$ is the area of a pixel [cm$^{2}$].

\section{Experiment}
\label{sec:experiment}
In this section, we provide an overview of the experimental setup including a parts list and associated costs, and the measurement procedure for the photon transfer curve, the QE transfer curve, and the transmission of filters or sensor windows. 

\subsection{System overview}
\begin{figure}[htb!]
\begin{center}
\begin{tabular}{c} 
\includegraphics[height=6cm]{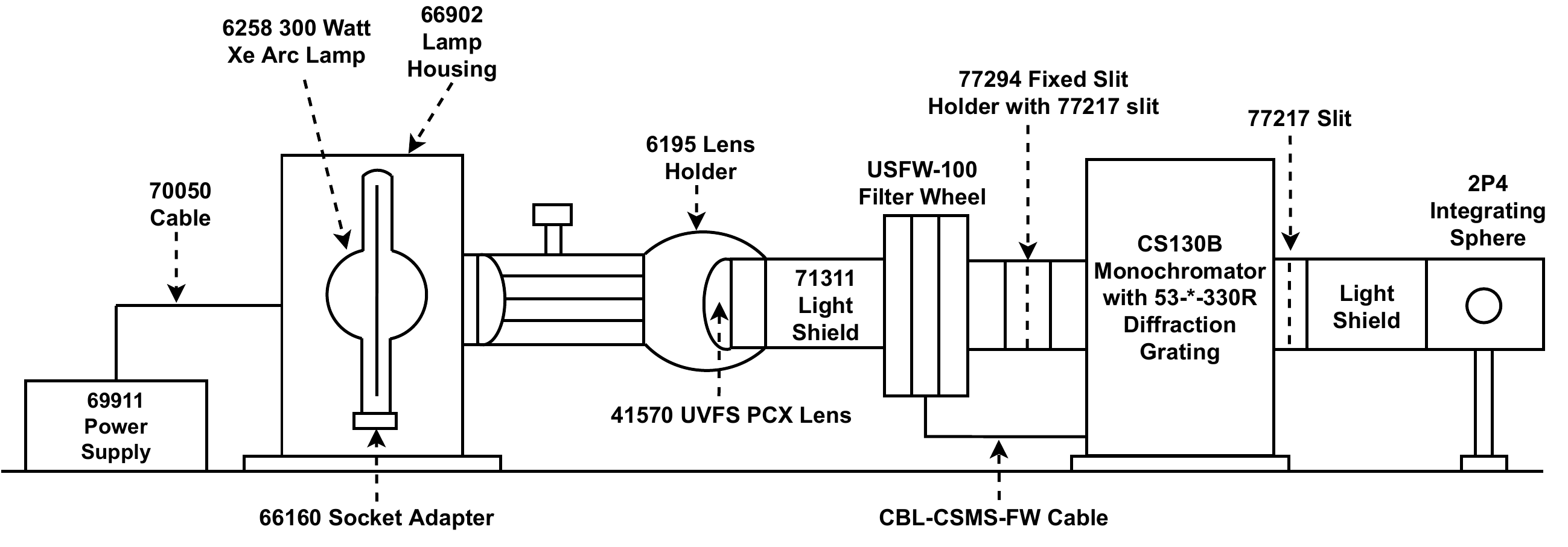}
\end{tabular}
\end{center}
\caption[example]
{\label{fig:block_diag}
  {Block diagram of the experiment.}}
\end{figure}

The block diagram of the experimental setup to construct the photon and QE transfer curves is shown in Figure \ref{fig:block_diag}. The parts list, the vendor, and associated costs are given in Table \ref{tab:parts}. We describe the key components below. 

\subsubsection{Lamp}
\begin{figure}[htb!]
\begin{center}
\begin{tabular}{c} 
\includegraphics[height=7cm]{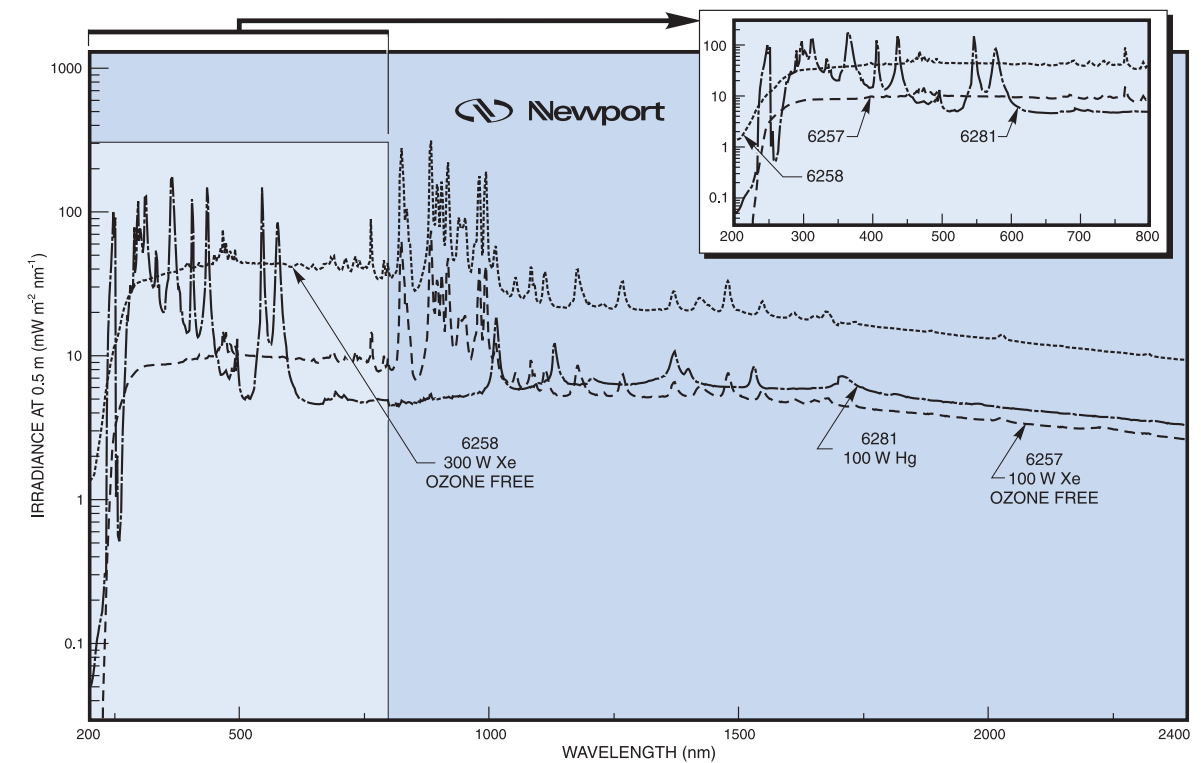}
\end{tabular}
\end{center}
\caption[example]
{\label{fig:spec_arc}
  {Spectral irradiance of the 6258 Xenon arc lamp.}}
\end{figure}

We used the 300 Watt 6258 Xenon Arc lamp, which has a broad spectrum from 200 to 2400 nm (see Figure \ref{fig:spec_arc}\footnote{https://www.newport.com/p/6258}). We used the ozone free version, as UV radiation below 242 nm produces toxic ozone.  The lamp has a lifetime of 900 hours. 

\subsubsection{Power supply}
The 69911 arc lamp power supply is highly regulated and provides a very stable output with a light ripple of $< 1\%$ RMS, accuracy of $<0.1\%$ of full scale, and line regulation of $\pm 0.05 \%$. The supply also has an RS-232 interface to control the lamp's operation parameters, monitor the output level, and turn it on/off remotely.

\subsubsection{Filter wheel}
The USFW-100 is a flange-mounted universal filter wheel capable of holding up to six 1-inch diameter filters. The filter wheel interfaces with the CS130B monochromator. For the QE transfer measurements, we used three longpass filters with cutoff wavelengths of 305 nm, 570 nm, and 1000 nm. 

\subsubsection{Monochromator}
\begin{figure}[htb!]
\begin{center}
\begin{tabular}{c} 
\includegraphics[height=7cm]{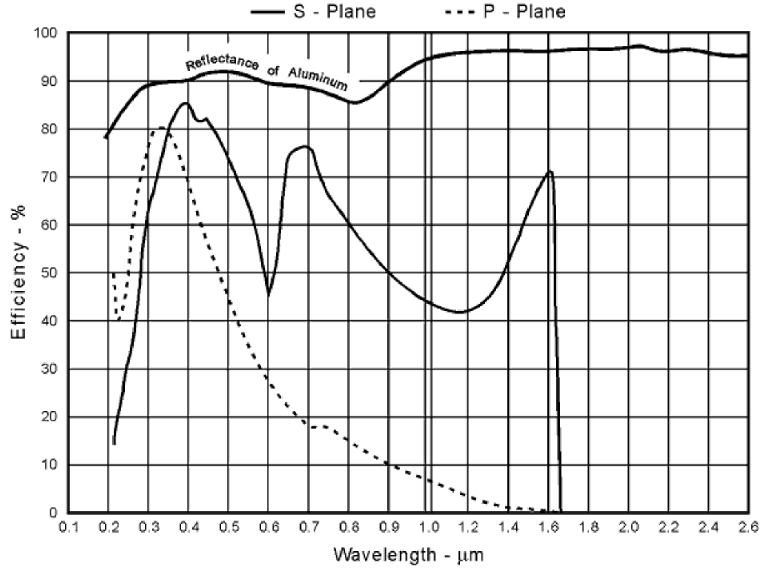}
\end{tabular}
\end{center}
\caption[example]
{\label{fig:grating_eff}
  {Efficiency of the 53-*-330R diffraction grating used for this work.}}
\end{figure}

The Oriel CS130B 1/8 meter monochromator is a low-cost, user-friendly instrument. The monochromator allows for monochromatric illumination of the sensor to measure the QE at different wavelengths. The CS130B has a stray light of 0.03$\%$, a wavelength accuracy of $\pm 0.25$ nm, and a wavelength precision of $\pm 0.0075$ nm. It provides motorized wavelength control with the Oriel Mono Utility software available on Windows and MacOS. It also has a USB and RS-232 computer interface for automated control. The monochromator supports up to 2 diffraction gratings. For this work, we used the 53-*-330R diffraction grating\footnote{https://www.newport.com/f/330r-plane-ruled-diffraction-gratings} from Newport (see Figure \ref{fig:grating_eff}).

\subsubsection{Calibrated photodiode}
\begin{figure}[htb!]
\begin{center}
\begin{tabular}{c} 
\includegraphics[height=6cm]{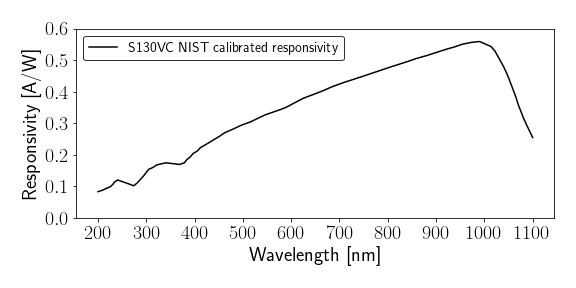}
\end{tabular}
\end{center}
\caption[]
{\label{fig:s130vc}
  {NIST calibrated responsitivity of the S130VC photodiode.}}
\end{figure}

We used the S130VC (UV-extended) silicon photodiode (200-1100 nm) from Thorlabs. The responsivity of the photodiode (see Figure \ref{fig:s130vc}) is calibrated by the National Insititute of Standards and Technology (NIST). The induced photocurrent can be read using the PM100USB power sensor and the Optical Power Monitor software.

\subsubsection{Integrating sphere}
\begin{figure}[htb!]
\begin{center}
\begin{tabular}{c} 
\includegraphics[height=6cm]{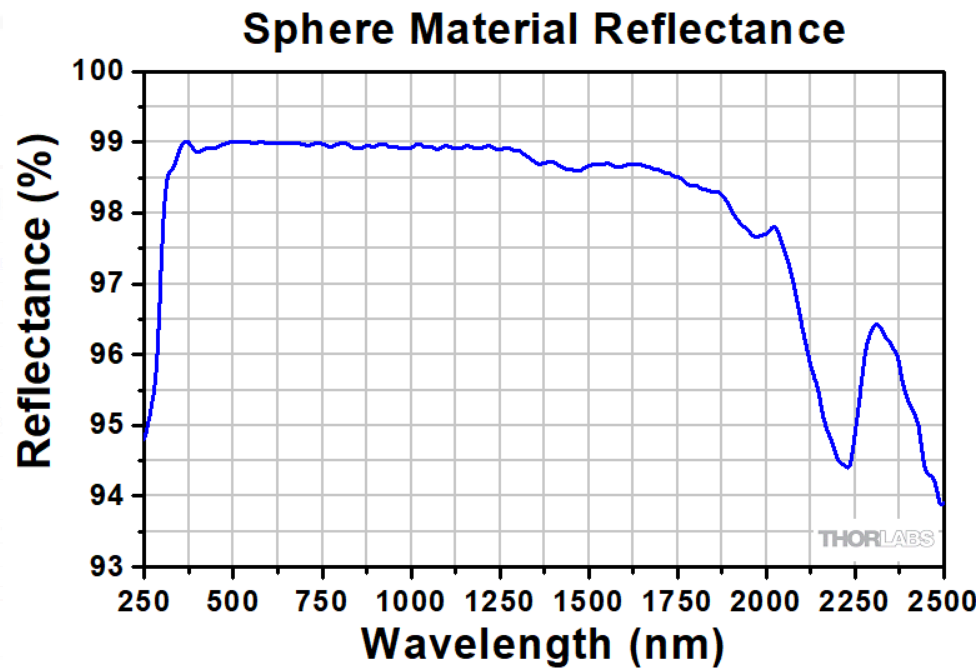}
\end{tabular}
\end{center}
\caption[]
{\label{fig:is200_R}
  {Reflectance of the IS200-4 integrating sphere.}}
\end{figure}

We used the IS200 4-port, 2-inch diameter integrating sphere from Thorlabs. The reflectance of the IS200 is shown in Figure \ref{fig:is200_R}. Note, this product is obsolete at the time of writing and seems to have been replaced by the 2P4 integrating sphere, which should work just as well.

\begin{table}[]
\begin{tabular}{|c|c|c|c|r|}
\hline
\textbf{Component}                         & \textbf{Part number} & \textbf{Vendor} & \textbf{Quantity} & \textbf{Cost (USD)} \\ \hline
300 Watt Xenon Arc Lamp (Ozone Free)             & 6258                 & Newport         & 1                 & 750.00              \\ \hline
Arc Lamp Housing                           & 66902                & Newport         & 1                 & 3,595.00             \\ \hline
Arc Lamp Power Supply                      & 69911                & Newport         & 1                 & 5,898.00             \\ \hline
Power Supply Cable                         & 70050                & Newport         & 1                 & 159.00              \\ \hline
Lamp Socket Adapter                        & 66160                & Newport         & 1                 & 115.00              \\ \hline
Flanged Lens Holder                        & 6195                 & Newport         & 1                 & 95.00               \\ \hline
Focusing Lens UVFS                         & 41570                & Newport         & 1                 & 235.00              \\ \hline
Adjustable Light Shield                    & 71311                & Newport         & 1                 & 178.00              \\ \hline
Universal Filter Wheel                     & USFW-100             & Newport         & 1                 & 1,719.00             \\ \hline
Fixed Slit                                 & 77217                & Newport         & 2                 & 390.00              \\ \hline
Fixed Slit Holder                          & 77294                & Newport         & 1                 & 239.00              \\ \hline
Filter Wheel Cable                         & CBL-CSMS-FW          & Newport         & 1                 & 80.00               \\ \hline
Filter Holder                              & LT10-05              & Newport         & 1                 & 17.50               \\ \hline
Spanner Wrench                             & LT05-WR              & Newport         & 1                 & 20.00               \\ \hline
Diffraction Grating                        & 53-*-330R            & Newport         & 1                 & 155.00              \\ \hline
Monochromator                              & CS130B               & Newport         & 1                 & 7,500.00             \\ \hline
Longpass Filter, 305 nm Cutoff             & 10CGA-305            & Newport         & 1                 & 50.00               \\ \hline
Longpass Filter, 570 nm Cutoff             & 10CGA-570            & Newport         & 1                 & 50.00               \\ \hline
Longpass Filter, 1000 nm Cutoff            & 10CGA-1000           & Newport         & 1                 & 50.00               \\ \hline
Integrating Sphere, 4 Port, 50 mm Diameter & 2P4                  & Thorlabs        & 1                 & 1,158.25             \\ \hline
Blackout Fabric                            & BK5                  & Thorlabs        & 2                 & 118.34              \\ \hline
UV Safety Laser Glasses                    & LG3                  & Thorlabs        & 1                 & 169.64              \\ \hline
Calibrated Photodiode Power Sensor         & S130VC               & Thorlabs        & 1                 & 669.94              \\ \hline
USB Power Meter Readout Interface          & PM100USB             & Thorlabs        & 1                 & 464.75              \\ \hline
\textbf{Total Cost}                        & \textbf{-}           & \textbf{-}      & \textbf{-}        & \textbf{\$23,876.42}   \\ \hline
\end{tabular}
\caption{List of components used for this work. All costs as of late 2021 - early 2022.}
\label{tab:parts}
\end{table}

\subsection{Procedure: photon transfer curve}
\label{sec:ptc_proc}
\begin{enumerate}
\item Construct the setup shown in Figure \ref{fig:block_diag}. See also Figure \ref{fig:exp1}.
\item Set the output of the monochromator to a single wavelength. We used 656 nm (H$\alpha$ line) for this work. 
\item Place the center of the sensor at the same height $h$ as the center of the output port (with reference to the optical bench). 
\item Place the sensor $\sim 10D$ away from the output port of the integrating sphere, where $D$ is the diameter of the output port, to uniformly illuminate the sensor.
\item Enclose the sensor and the integrating sphere with a cardboard box to minimize background light (see Figure \ref{fig:exp2}). 
\item Place the BK5 blackout material over the box to further minimize background light (see Figure \ref{fig:exp3}).
\item Turn on the power supply to the arc lamp. Turn on the filter wheel and the monochromator. Wait at least 15 minutes for the lamp to ignite and stabilize. 
\item We need to acquire a series of light images at different exposure times, starting with the shortest possible exposure time (typically $\sim$ 1 ms for mechanical shutters) and increasing exposure time until saturation. At each exposure level, take at least three light frames and three dark frames (with the monochromator shutter off), which we will average for better estimates of the signal and the noise. 
\item After the data has been collected, construct two individual datasets: (i) total noise data and (ii) shot and read noise data. 
\item Total noise data
\begin{itemize}
    \item At each exposure level, median stack (pixel-by-pixel) the dark frames to create a master dark frame. 
    \item Subtract the master dark frame from a master light frame (similarly constructed as the master dark frame) to construct a ``clean" frame.
    \item Dark frame subtraction removes an average offset level (an average of the camera's output in ADUs in the absence of signal electrons) as well as the dark current. Note, it is not strictly necessary to subtract the dark current from the signal, as it is not important \textit{how} the signal is generated for the PTC, as long as the source exhibits shot noise characteristics, which dark current does. 
    \item Hence, a PTC can be constructed from dark current measurements only, and a light source is not required. However, since pixel-to-pixel dark current non-uniformity is typically larger than pixel-to-pixel sensitivity non-uniformity, we recommend using a light source and subtracting the dark current. \item At each exposure level, the signal and the noise estimates in ADUs are the average and the standard deviation of the clean frame.
\end{itemize}
\item Shot and read noise data
\begin{itemize}
    \item To estimate the read noise and the camera gain $K$, the FPN must be removed to obtain a shot plus read noise only curve. 
    \item At each exposure level, the FPN can be removed by pixel-by-pixel subtracting two light frames taken back-to-back at the same exposure time, with correction for the increase in random noise due to frame subtraction.
    
    \begin{equation}
        \sigma_{\mathrm{READ+SHOT}} \,[\mathrm{ADU}] = \left[\frac{\sum_{i=1}^{N_{\mathrm{pix}}}(x_{i} - y_{i})^{2}}{2 N_{\mathrm{pix}}} \right]^{1/2}
    \end{equation}
\noindent where $x_{i}$ and $y_{i}$ are the signal values at the $i$th pixel of the first and second frame, respectively.
\end{itemize}
\item After the total and shot plus read noise data has been collected, plot the two curves on a logarithmic scale.
\item Gain: Fit equation \ref{eqn:shot_noise} to the read noise plus shot noise data to get the gain constant $K$.
\item Read noise: The constant offset at low exposure levels for the read noise plus shot noise data is the read noise in ADUs. Multiply by $K$ to convert it to electrons per pixel.  
\item Fixed pattern noise: Fit equation \ref{eqn:fpn} to the total noise data in the FPN regime to measure the fixed pattern noise quality factor $P_{\mathrm{N}}$. 
\item Full well capacity: The regime when noise decreases and saturation occurs provides the pixel full-well capacity.
\item Dynamic range can be estimated (in decibels) by the ratio of the full well and the read noise. 

\begin{equation}
\mathrm{DR} = 20 \log{ \left(\frac{S_{\mathrm{FW}}\,[e^{-}]}{\sigma_{\mathrm{R}}\,[e^{-}]} \right)}
\end{equation}
\end{enumerate}

\subsection{Procedure: quantum efficiency transfer curve}
\begin{enumerate}
\item Recall from equation \ref{eqn:qe_eqn} that the QE depends on the sensor term and the photodiode term. 
\item Sensor term
\begin{itemize}
\item We will need the count rates at different wavelengths, the camera gain constant, and the pixel area. 
\item Repeat steps 1, 3, 4, 5, 6, and 7 given in Section \ref{sec:ptc_proc}. Note that for step 4, for measurements below 250 nm, the light intensity at the output may be too low to generate a signal count rate above the noise level. In this case, we suggest placing the sensor as close as possible to the output port of the integrating sphere.
\item Take multiple (we recommend at least 5) background frames (with the monochromator shutter on) at various exposure times (we recommend 1 sec, 5 sec, and 10 sec, but it might be necessary to expose for longer).
\item We recommend using longpass filters in the filter wheel during data collection to minimize potential light leakage from other wavelengths. The onset wavelengths of the filters we used were 305 nm, 570 nm, and 1000 nm.
\item Turn the monochromator shutter off. Collect light frames at the different wavelengths of interest. For each wavelength, take multiple frames of the same exposure time and repeat for at least a couple of different exposure times (while ensuring you have background frames of the same exposures). 
\item After the data has been collected, turn the monochromator shutter on and remove the camera. Do not turn off the lamp.
\item At each wavelength, subtract the master background frame (median stack) from the master light frame (median stack) to get the clean frame. The signal level (in ADUs) and its uncertainty is then the average and standard deviation of the clean frame. Use the gain $K$ (measured from the PTC) and the exposure time to get a signal count rate in units of electrons/second/pixel. 
\end{itemize}
\item Photodiode term
\begin{itemize}
    \item We will need the photocurrent at different wavelengths, the background current, the responsivity, and the active area of the photodiode. 
    \item Replace the sensor with the photodiode, ensuring that the photodiode is at the same height and distance away from the output port of the integrating sphere as the sensor (see Figure \ref{fig:exp4}).
    \item Enclose the photodiode and the integrating sphere in a cardboard box to minimize background light (see Figure \ref{fig:exp2}).
\item Place the BK5 blackout material over the box to further minimize background light (see Figure \ref{fig:exp3}). 
\item Estimate the background photodiode current. The current can be read using the Optical Power Monitor software and the PM100USB power sensor. We recommend collecting the current data over a period of at least 15 minutes and taking the average and standard deviation of the data as the background current estimate.
\item Turn the monochromator shutter off. Now, we are ready to collect the photocurrent data over different wavelengths. At each wavelength, record the current data for at least 5 minutes and calculate the mean and standard deviation for the photocurrent estimate. Use the longpass filters at the same wavelengths as the sensor data collection step. 
\end{itemize}
\end{enumerate}

\subsection{Procedure: window or filter transmission measurement}
It is typical for COTS cameras to have a window over the sensor. To get a more accurate QE measurement of the sensor, it is necessary to measure the transmission of the window independently, to quantify the loss in transmission due to the window itself. We provide reference to an instrument that we used to measure the transmission of camera windows and UV/optical filters. We used the Lambda 365 UV/Vis Spectrophotometer\footnote{https://www.perkinelmer.com/product/lambda-365-spectrophotometer-uv-express-n4100020} and a film holder. The Lambda 365 is both accurate and precise and can measure the transmission from 190 to 1100 nm with a spectral resolution of 0.5 nm. 

\section{Results} \label{sec:results}
\begin{figure}[htb!]
\begin{center}
\begin{tabular}{c} 
\includegraphics[height=9cm]{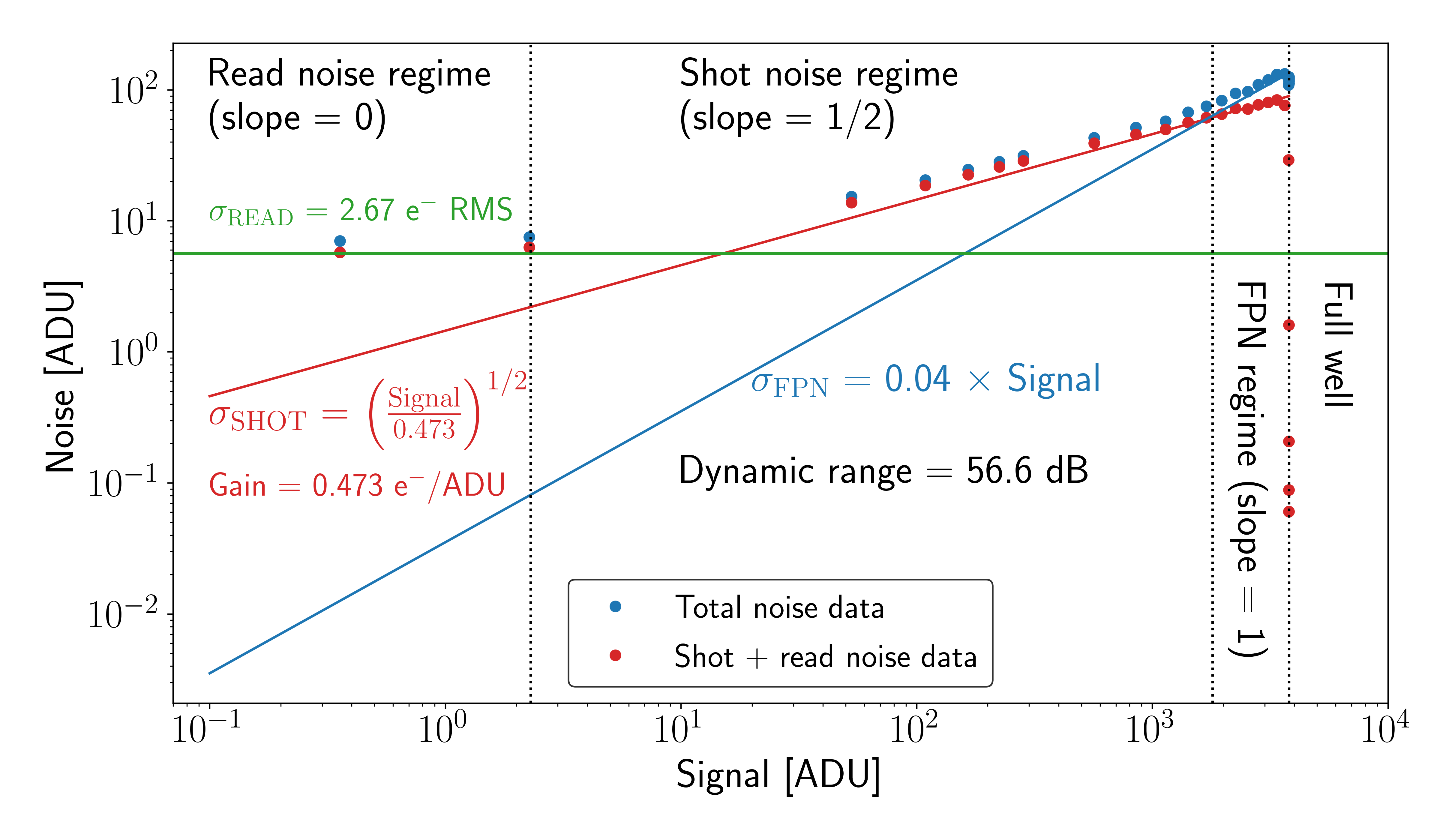}
\end{tabular}
\end{center}
\caption[]
{\label{fig:ptc_2020}
  {Photon transfer curve for the Aluma 2020 BSI (GSENSE 2020 BSI CMOS sensor).}}
\end{figure}

\begin{figure}[htb!]
\begin{center}
\begin{tabular}{c} 
\includegraphics[height=9cm]{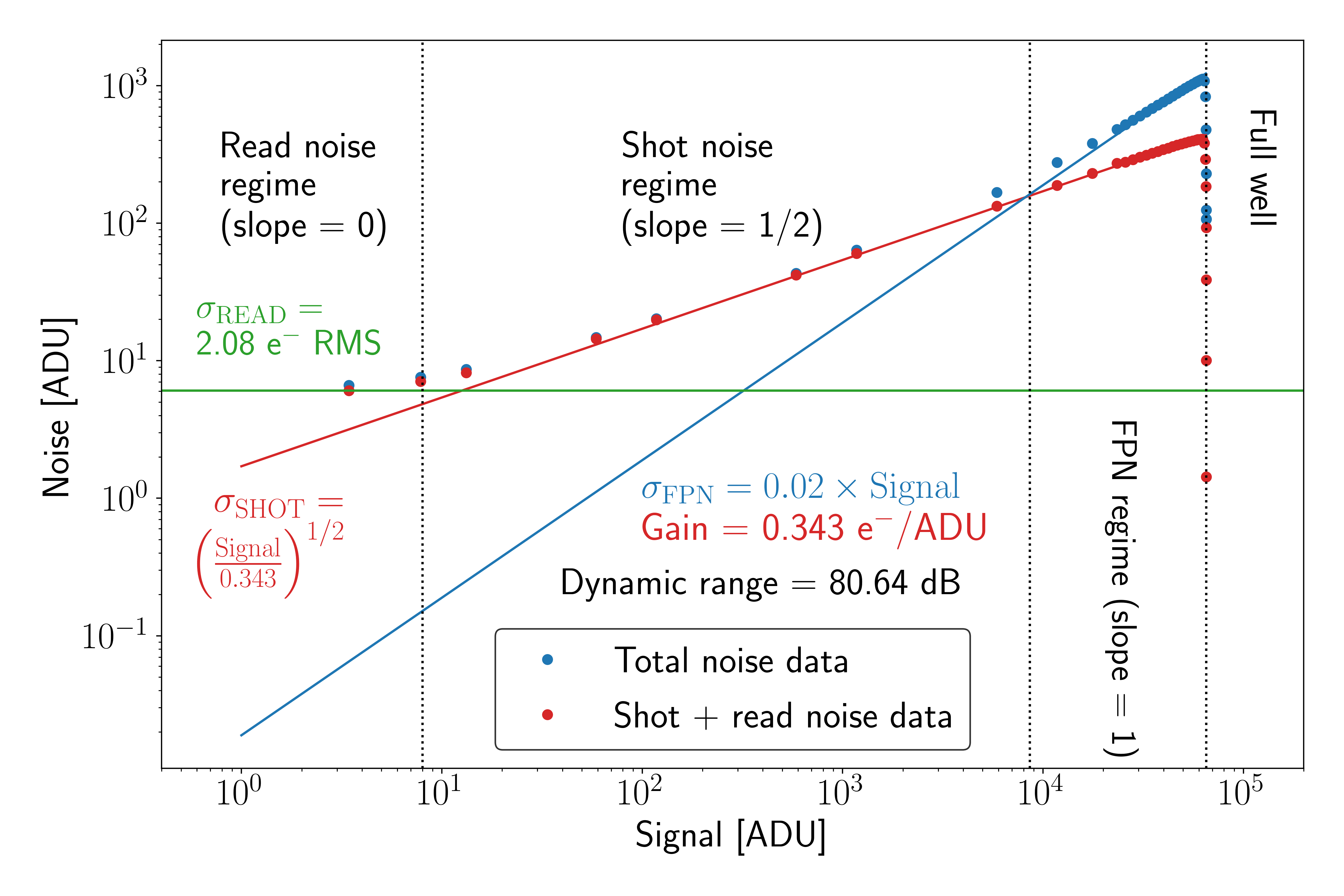}
\end{tabular}
\end{center}
\caption[]
{\label{fig:ptc_1hy}
  {Photon transfer curve for the QHY600 (Sony IMX 455 BSI CMOS sensor).}}
\end{figure}

\begin{figure}[htb!]
\begin{center}
\begin{tabular}{c} 
\includegraphics[height=8.5cm]{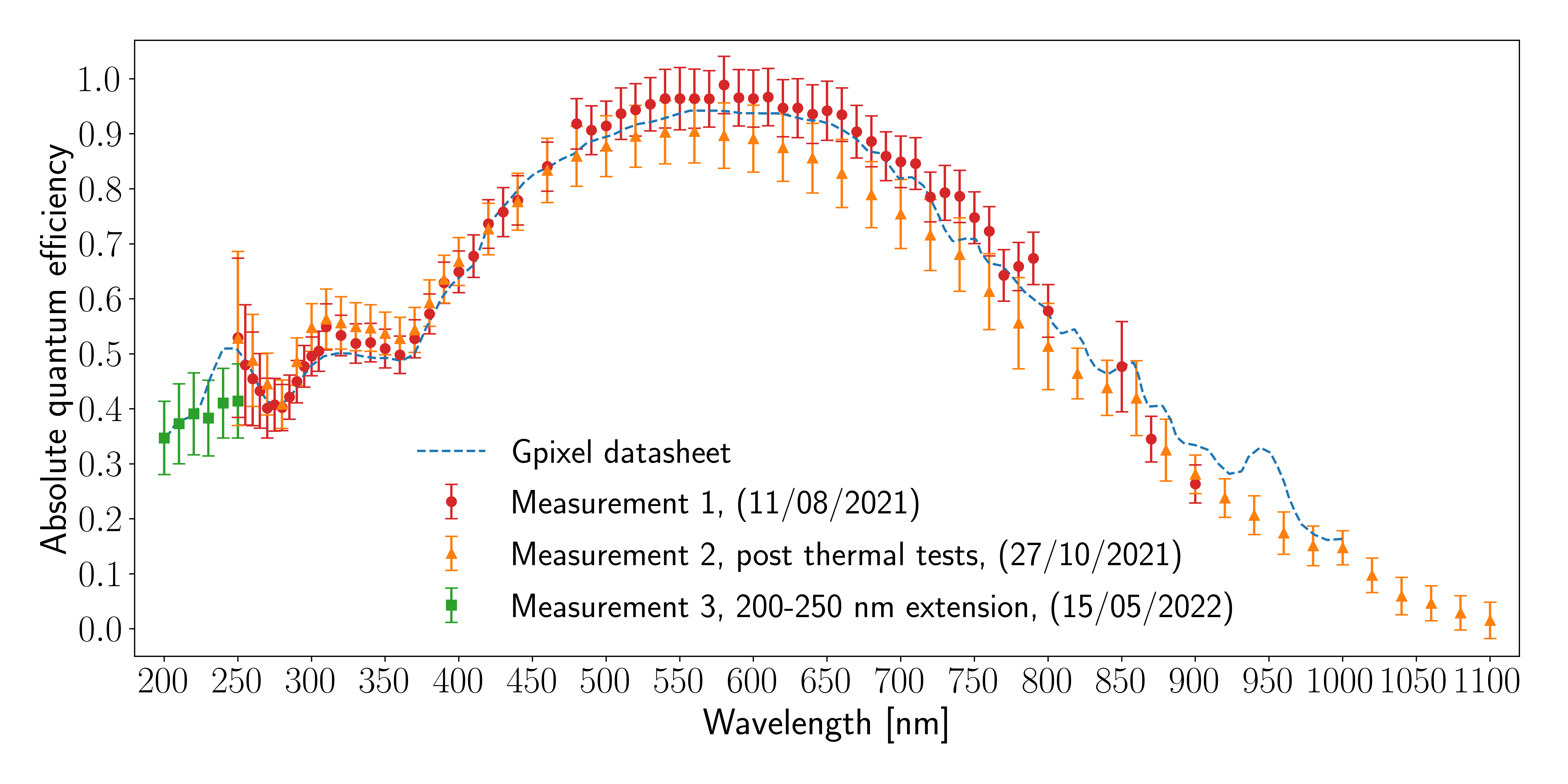}
\end{tabular}
\end{center}
\caption[]
{\label{fig:qe_2020}
  {Absolute quantum efficiency curve of the GSENSE2020 BSI CMOS sensor (with the Aluma 2020 BSI camera).}}
\end{figure}

\begin{figure}[htb!]
\begin{center}
\begin{tabular}{c} 
\includegraphics[height=8.5cm]{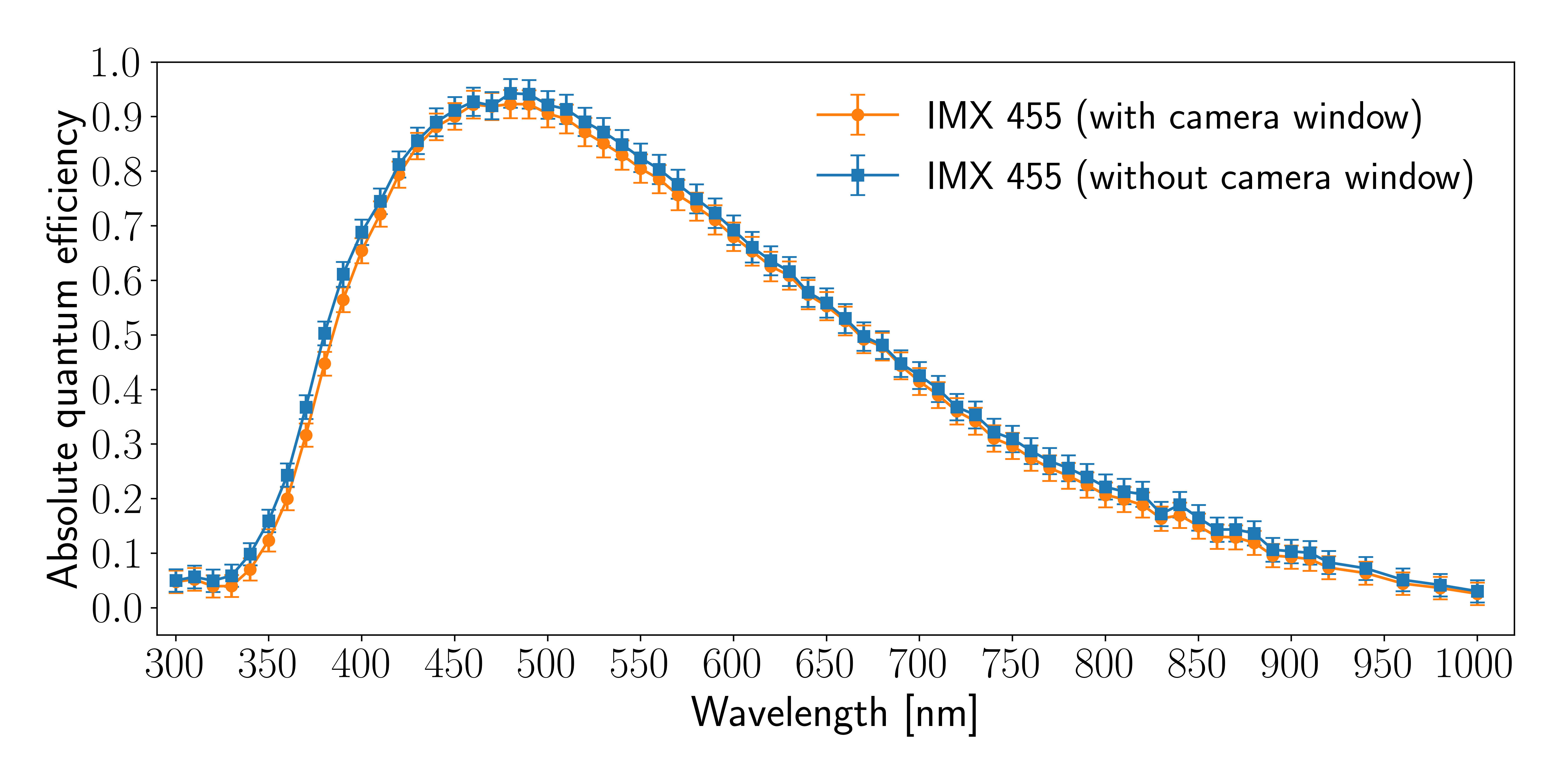}
\end{tabular}
\end{center}
\caption[]
{\label{fig:qe_455}
  {Absolute quantum efficiency curve of the Sony IMX455 BSI CMOS sensor (with the QHY600 camera).}}
\end{figure}

The photon transfer curves of the Aluma 2020 BSI (GSENSE 2020 BSI CMOS sensor) and the QHY600 (Sony IMX 455 CMOS sensor) are shown in Figures \ref{fig:ptc_2020} and \ref{fig:ptc_1hy}, respectively. The read noise, gain, FPN, the dynamic range, and the full well capacity are highlighted. The data for the Aluma 2020 BSI was collected in the ``High" gain mode. For the QHY600, a ``Gain Setting" of 56 was used. The QE transfer curve for the GSENSE 2020 BSI CMOS sensor is shown in Figure \ref{fig:qe_2020}. We measured the GSENSE 2020 BSI QE transfer curve on three different occasions: (i) measurement 1 (initial measurement), (ii) measurement 2 (post thermal testing), and (iii) measurement 3 (200-250 nm extension). The QE transfer curve for the Sony IMX 455 CMOS BSI sensor is shown in Figure \ref{fig:qe_455}, both with and without the camera window. We estimated the systematic uncertainty to be $\sim$ 2$\%$ by collecting data for the sensor and photodiode term four different iterations at a single wavelength, where we disassembled and reassembled the entire experimental setup from scratch at each iteration.

\section{Conclusion}
This paper presents a low-cost method for measuring important parameters of  CCD/CMOS sensors, including the read noise, the gain, and the absolute quantum efficiency. In particular, it is vital to understand the absolute quantum efficiency as a function of wavelength for sensors on astronomical instruments. Since the quantum efficiency can set the sensitivity limit for the detection of sources, measuring the quantum efficiency informs the observation planning and the scientific capability of the instrument. We present the experimental setup (including a parts list and figures) and a step-by-step procedure for constructing both the photon and the quantum efficiency transfer curves, with the hope to be useful for academic and industry institutions looking to build a similar setup to characterize their own CCD/CMOS sensors. Finally, we also present the results of the photon and quantum efficiency transfer curves of two commercial-off-the-shelf sensors, the GSENSE 2020 BSI CMOS sensor and the Sony IMX 455 BSI CMOS sensor. 

\section{Acknowledgements}
We acknowledge the camera vendors, Diffraction Limited and QHYCCD  for their technical support. We thank the Analytical Laboratory for Environmental Science Research and Training (ANALEST) facility at the University of Toronto for using the Lambda 365 UV-Vis Spectrophotometer. We also thank the seed funding from the Dunlap Institute for Astronomy and Astrophysics at the University of Toronto that helped enable this work. 
\bibliography{ref} 

\begin{thebibliography}{1}

\bibitem{Jacquot_2011}
B.~C. Jacquot, S.~P. Monacos, M.~E. Hoenk, F.~Greer, T.~J. Jones, and
  S.~Nikzad, ``A system and methodologies for absolute quantum efficiency
  measurements from the vacuum ultraviolet through the near infrared,'' {\em
  Review of Scientific Instruments}~{\bf 82}, p.~043102, Apr 2011.

\bibitem{Sperlich_2013}
K.~Sperlich and H.~Stolz, ``Quantum efficiency measurements of ({EM}){CCD}
  cameras: high spectral resolution and temperature dependence,'' {\em
  Measurement Science and Technology}~{\bf 25}, p.~015502, Nov 2013.

\bibitem{Coles_2017}
R.~Coles, J.~Chiang, D.~Cinabro, J.~Haupt, H.~Neal, A.~Nomerotski, and
  P.~Takacs, ``An automated system to measure the quantum efficiency of {CCDs}
  for astronomy,'' {\em Journal of Instrumentation}~{\bf 12},
  pp.~C04014--C04014, Apr 2017.

\bibitem{Krishnamurthy_2017}
A.~Krishnamurthy, J.~Villasenor, S.~Kissel, G.~Ricker, and R.~Vanderspek, ``An
  optical test bench for the precision characterization of absolute quantum
  efficiency for the {TESS} {CCD} detectors,'' {\em Journal of
  Instrumentation}~{\bf 12}, pp.~C05013--C05013, May 2017.

\bibitem{Bastian_Querner_2021}
B.~Bastian-Querner, N.~Kaipachery, D.~Küster, J.~Schliwinski, S.~Alfassi,
  A.~Asif, M.~F. Barschke, S.~Ben-Ami, D.~Berge, A.~Birman, R.~Bühler, N.~D.
  Simone, A.~Fenigstein, A.~Gal-Yam, G.~Giavitto, J.~H. Crespo, D.~Ivanov,
  O.~Katz, M.~Kowalski, S.~Kulkarni, O.~Lapid, T.~Liran, E.~Netzer, E.~O. Ofek,
  S.~Philipp, H.~Prokoph, S.~Regev, Y.~Shvartzvald, M.~Vasilev, D.~Veinger,
  J.~Watson, E.~Waxman, S.~Worm, and F.~Zappon, ``Sensor characterization for
  the {ULTRASAT} space telescope,'' in {\em {UV}/Optical/{IR} Space Telescopes
  and Instruments: Innovative Technologies and Concepts X},  J.~B.
  Breckinridge, H.~P. Stahl, and A.~A. Barto, eds., {SPIE}, Aug 2021.

\bibitem{Law_2022}
N.~M. Law, H.~Corbett, N.~W. Galliher, R.~Gonzalez, A.~Vasquez, G.~Walters,
  L.~Machia, J.~Ratzloff, K.~Ackley, C.~Bizon, C.~Clemens, S.~Cox,
  S.~Eikenberry, W.~S. Howard, A.~Glazier, A.~W. Mann, R.~Quimby, D.~Reichart,
  and D.~Trilling, ``Low-cost access to the deep, high-cadence sky: the argus
  optical array,'' {\em Publications of the Astronomical Society of the
  Pacific}~{\bf 134}, p.~035003, Mar 2022.

\bibitem{Romualdez_2020}
L.~J. Romualdez, S.~J. Benton, A.~M. Brown, P.~Clark, C.~J. Damaren, T.~Eifler,
  A.~A. Fraisse, M.~N. Galloway, A.~Gill, J.~W. Hartley, B.~Holder, E.~M. Huff,
  M.~Jauzac, W.~C. Jones, D.~Lagattuta, J.~S.-Y. Leung, L.~Li, T.~V.~T. Luu,
  R.~J. Massey, J.~McCleary, J.~Mullaney, J.~M. Nagy, C.~B. Netterfield,
  S.~Redmond, J.~D. Rhodes, J.~Schmoll, M.~M. Shaaban, E.~Sirks, and S.-I. Tam,
  ``Robust diffraction-limited near-infrared-to-near-ultraviolet wide-field
  imaging from stratospheric balloon-borne platforms{\textemdash}super-pressure
  balloon-borne imaging telescope performance,'' {\em Review of Scientific
  Instruments}~{\bf 91}, p.~034501, Mar 2020.

\bibitem{s_ccd}
J.~R. Janesick, {\em Scientific Charge-Coupled Devices}, SPIE—The
  International Society for Optical Engineering, Washington, 2001.

\bibitem{s_ccd_2}
J.~R. Janesick, {\em Photon Transfer: DN to $\lambda$}, The Society of
  Photo-Optical Instrumentation Engineers, Washington, 2007.

\end{thebibliography}
\bibliographystyle{spiebib} 

\newpage
\appendix
\section{Experimental setup}

\begin{figure}[htb!]
\begin{center}
\begin{tabular}{c} 
\includegraphics[height=9.5cm]{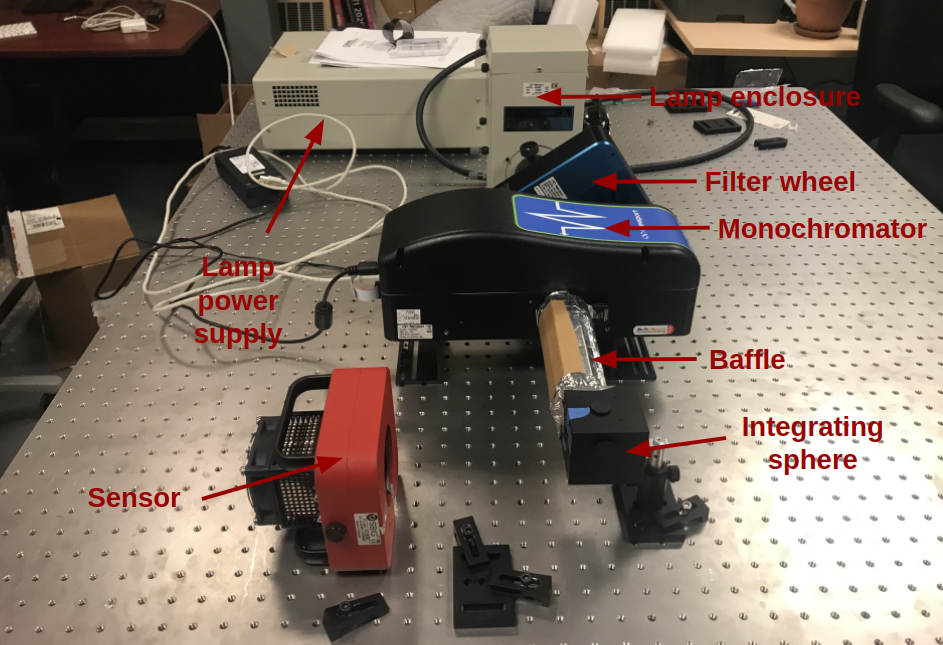}
\end{tabular}
\end{center}
\caption[]
{\label{fig:exp1}
  {Setup for sensor data collection.}}
\end{figure}

\begin{figure}[htb!]
\begin{center}
\begin{tabular}{c} 
\includegraphics[height=9.5cm]{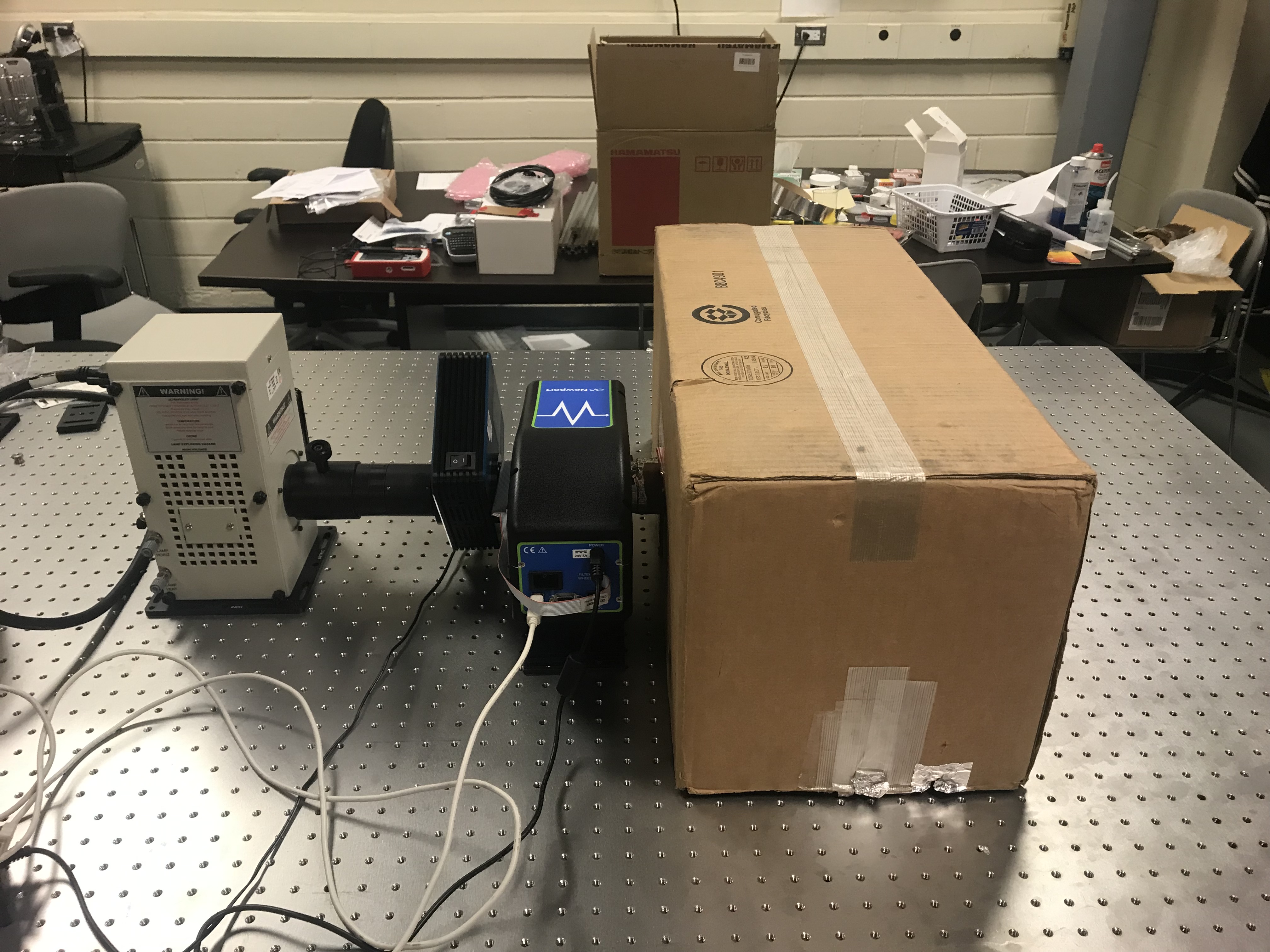}
\end{tabular}
\end{center}
\caption[]
{\label{fig:exp2}
  {Setup for sensor data collection: stray light minimization (stage 1).}}
\end{figure}

\begin{figure}[htb!]
\begin{center}
\begin{tabular}{c} 
\includegraphics[height=10cm]{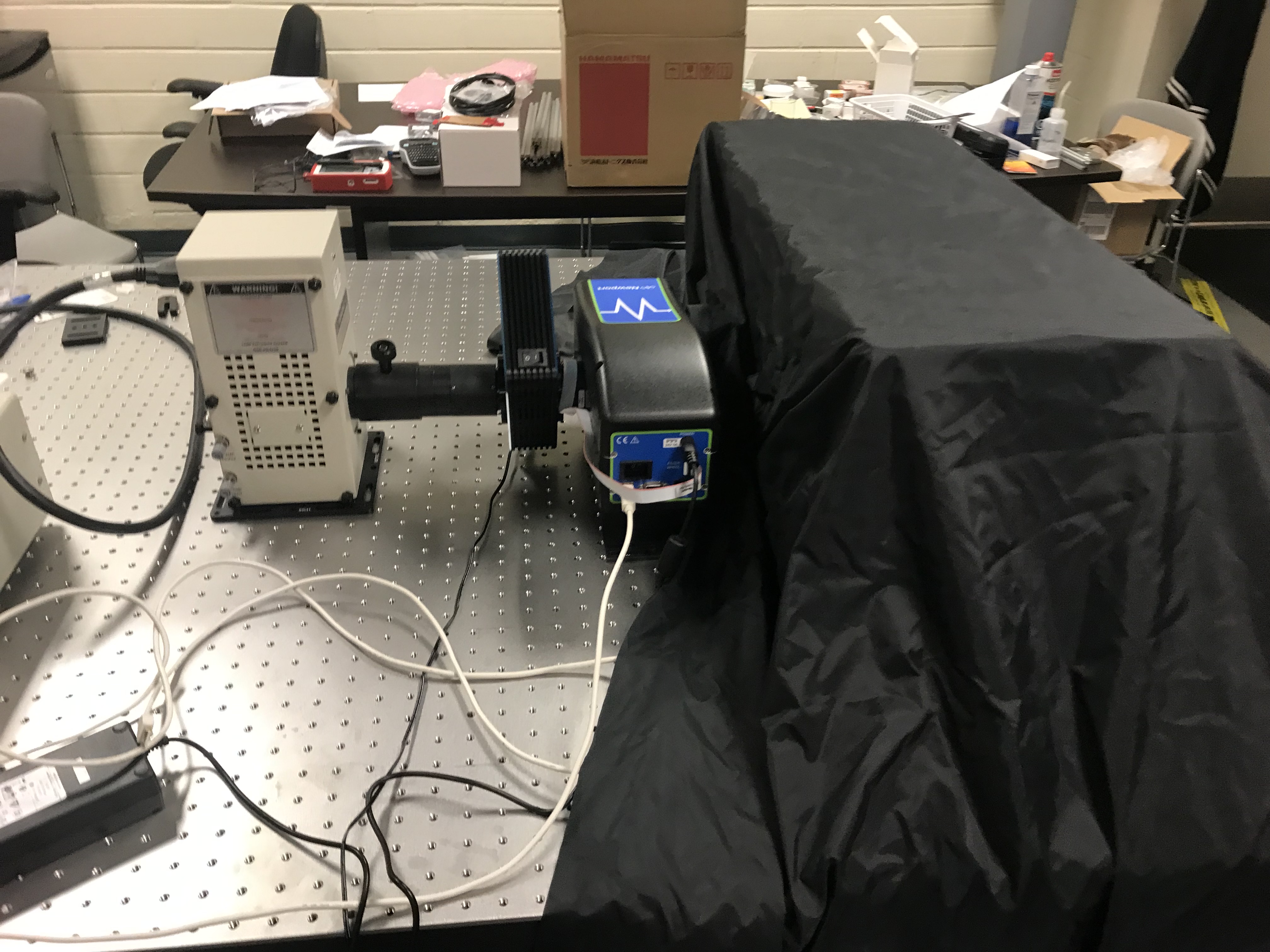}
\end{tabular}
\end{center}
\caption[]
{\label{fig:exp3}
  {Setup for sensor data collection: stray light minimization (stage 2).}}
\end{figure}

\begin{figure}[htb!]
\begin{center}
\begin{tabular}{c} 
\includegraphics[height=10cm]{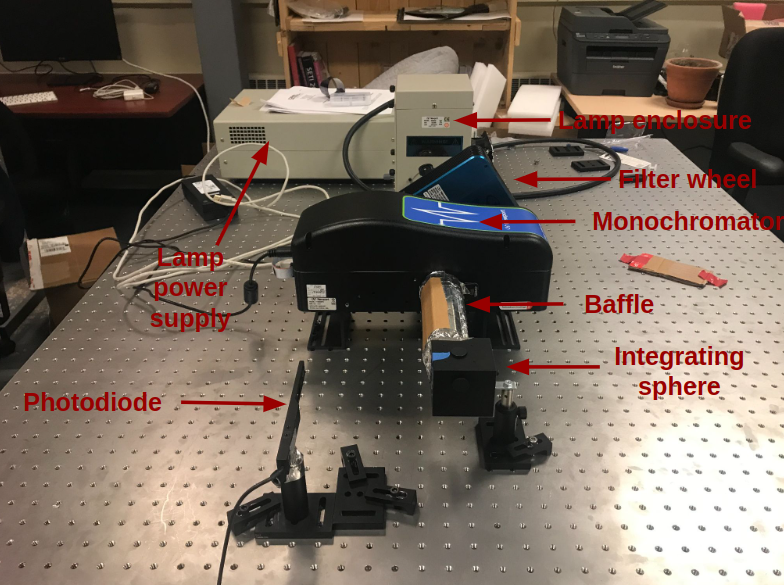}
\end{tabular}
\end{center}
\caption[]
{\label{fig:exp4}
  {Setup for photodiode data collection.}}
\end{figure}

\end{document}